\documentclass[]{spie}  %>>> use for US letter paper
\usepackage{graphicx}

\usepackage{soul,color}
\usepackage{array}
\usepackage{subfig}
\usepackage{floatrow}
\newfloatcommand{FigandTab}{table}[][\FBwidth]

\title{Analytic and SPICE modeling of stochastic ReRAM circuits}

\author{Vincent~J.~Dowling\supit{a}, Valeriy~A.~Slipko\supit{b} and Yuriy~V.~Pershin\supit{a}
\skiplinehalf
\supit{a}Department of Physics and Astronomy, University of South Carolina, Columbia, SC 29208, USA; \\
\supit{b}Institute of Physics, Opole University, Opole 45-052, Poland
}

\authorinfo{Further author information: (Send correspondence to Y.V.P.)\\Y.V.P: E-mail: pershin@physics.sc.edu, Telephone: +1 803 777 5073}
\begin{document}
  \maketitle

%%%%%%%%%%%%%%%%%%%%%%%%%%%%%%%%%%%%%%%%%%%%%%%%%%%%%%%%%%%%%
\begin{abstract}
The modeling of conventional (deterministic) electronic circuits -- ones consisting of transistors, resistors, capacitors, inductors, and other traditional electronic components -- is a well-established subject. The cycle-to-cycle variability of emerging electronic devices, in particular, certain ReRAM cells, has led to the concept of stochastic circuits. Unfortunately, even in relatively simple cases, the direct transient analysis of  stochastic circuits is computationally demanding and potentially impractical, if possible at all. An important development in this area has been the application of a master equation that is easily implemented in SPICE. In this conference paper, we briefly review the master equation approach and present an improved implementation of this approach in SPICE. Moreover, we find an attractor state in a periodically driven memristive circuit -- a stochastic counterpart of deterministic memristor attractors.
\end{abstract}

%>>>> Include a list of keywords after the abstract


%%%%%%%%%%%%%%%%%%%%%%%%%%%%%%%%%%%%%%%%%%%%%%%%%%%%%%%%%%%%%
\section{INTRODUCTION}
\label{sec:intro}  % \label{} allows reference to this section

It is well known and documented in many papers that resistance-switching memory cells -- such as electrochemical metallization (ECM) cells~\cite{valov2011electrochemical} and valence change memory (VCM) cells~\cite{lim2015conduction} -- exhibit significant device-to-device and cycle-to-cycle variability. More specifically, past experiments~\cite{jo2009programmable,gaba2013stochastic,naous2021theory} with ECM cells have shown that the probability of switching from their {off} (0) to {on} (1) state and back can be described by switching rates
 \begin{eqnarray}
 \gamma_{0\rightarrow 1}(V) &=& \left\{ \begin{array}{ccl}
\left( \tau_{01} e^{-V/V_{01}}\right)^{-1}&,& V>0   \\
0 &,& \textnormal{otherwise}
\end{array}\right. \;\; , \\
 \gamma_{1\rightarrow 0}(V) &=& \left\{ \begin{array}{ccl}
\left( \tau_{10} e^{-|V|/V_{10}}\right)^{-1}&,& V<0 \\
0 &,& \textnormal{otherwise}
\end{array}\right. \; , \label{eq:gamma}
\end{eqnarray}
where $V$ is the voltage across the device, and $\tau_{01(10)}$ and $V_{01(10)}$ are device-specific parameters. When subjected to a constant voltage, the switching time of such stochastic memristive devices follows a Poisson distribution~\cite{gaba2013stochastic}.

It is difficult to simulate electronic circuits with stochastic components as the behavior of such circuits, in general, is not deterministic. Importantly, the circuit variables -- such as voltages, currents, etc. -- can only be predicted on average and their distributions are described with the help of discrete probabilities or probability distribution functions. Recently, some notable advancements in the description of stochastic circuits have been reported. We have pioneered the use of a master equation for the description of electronic circuits with certain stochastic memristors~\cite{dowling2020probabilistic}, and the Kolmogorov–Chapman equation for circuits combining memristors with reactive components (capacitors and/or inductors)~\cite{Slipko21a}. These approaches can be readily applied to circuits with certain binary and multi-state memristors~\cite{dowling2020SPICE,Ntinas2020}. Moreover, we have shown how to implement the modeling of some probabilistic electronic circuits in SPICE~\cite{dowling2020SPICE} -- a general-purpose, open-source analog electronic circuit simulator~\cite{kundert2006designer,vladimirescu1994spice}.

In this paper, we give a brief introduction to the master equation approach followed by an interesting example of probabilistic circuit modeling. Previously, our focus was on dc-biased memristive circuits~\cite{Slipko21a,dowling2020SPICE} and results for the switching of in-parallel (Fig.~\ref{fig:1}(a)) and in-series connected probabilistic memristors were found. Here, we consider the circuit shown in Fig.~\ref{fig:1}(b) driven by an alternating polarity voltage. An interesting observation is the transition into a steady state, which is reminiscent of the attractor state for deterministic memristor circuits~\cite{Pershin2019a}. We emphasize that the SPICE implementation of Fig.~\ref{fig:1}(b) uses a more optimized approach.

%%%%%%%%%%%%%%%%%%%%%%%%%%%%%%%%%%%%%%%%%%%%%%%%%%%%%%%%%%%%%
\section{MASTER-EQUATION APPROACH}

%\subsection{Master equation}

%Following Ref.~\citeonline{dowling2020probabilistic},

\begin{figure}[tb]
  \centering
     (a)\includegraphics[width=0.12\columnwidth]{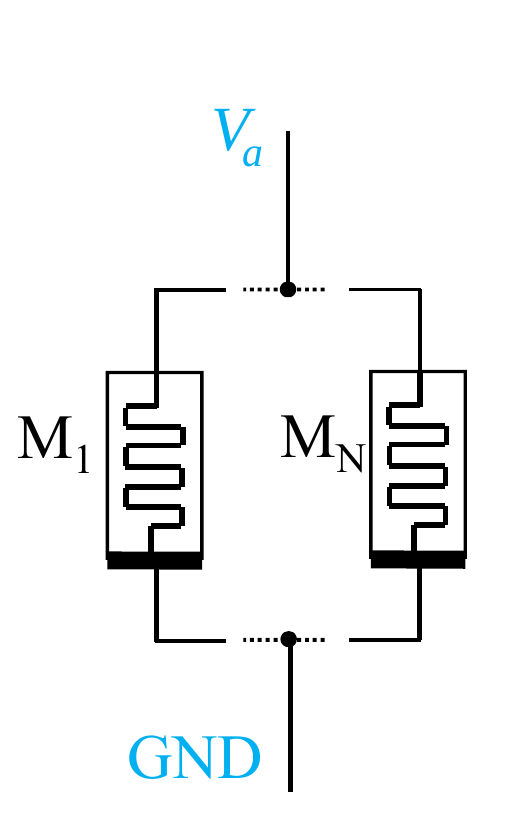} $\;$
     (b)\includegraphics[width=0.25\columnwidth]{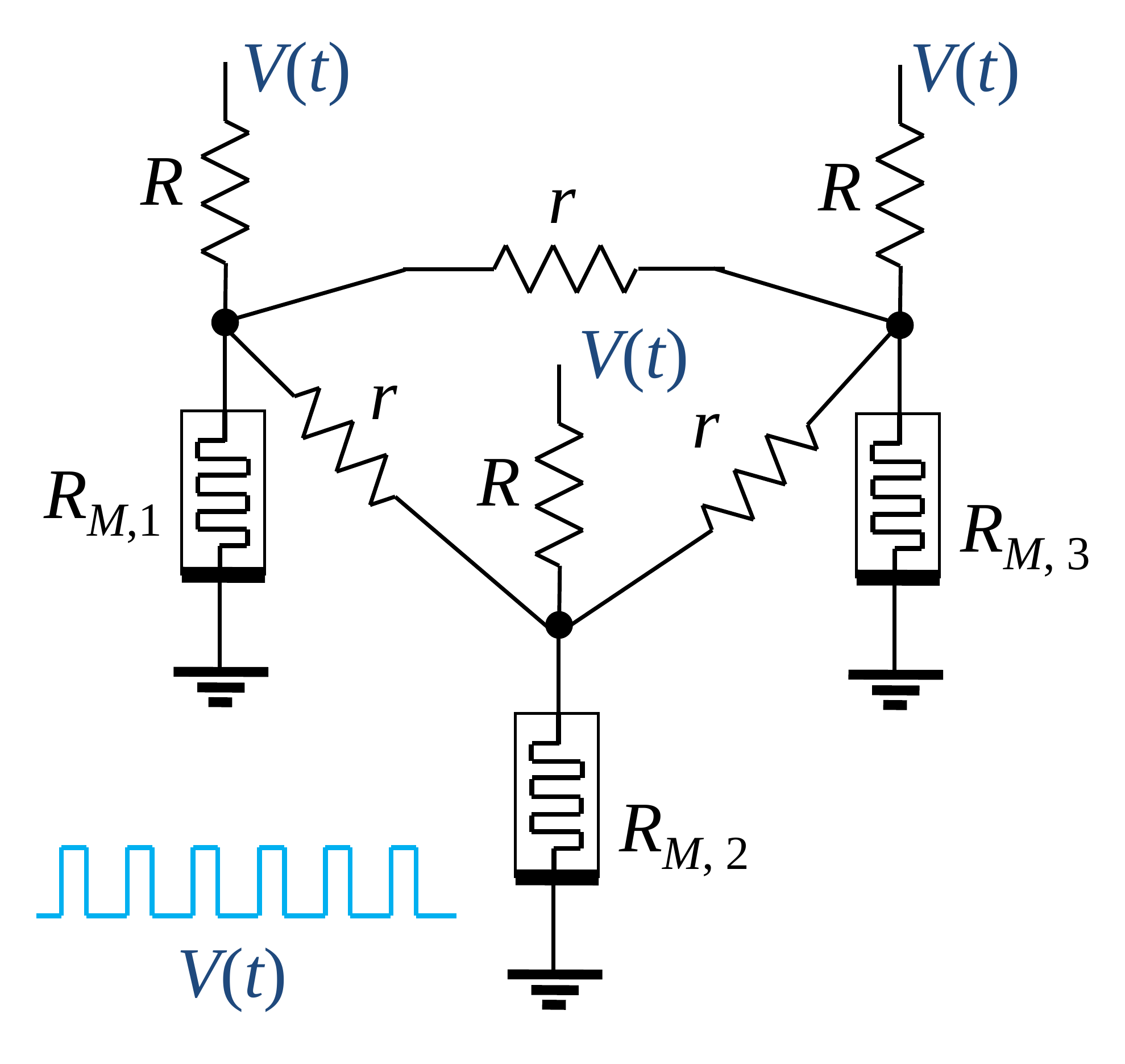} $\;$
     (c)\includegraphics[height=0.2\columnwidth]{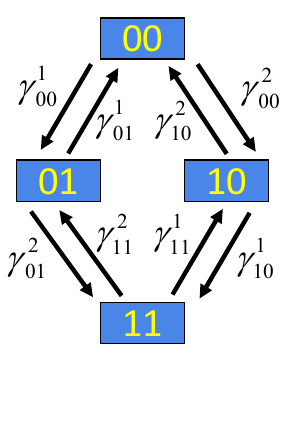} $\;$
     (d)\includegraphics[width=0.25\textwidth]{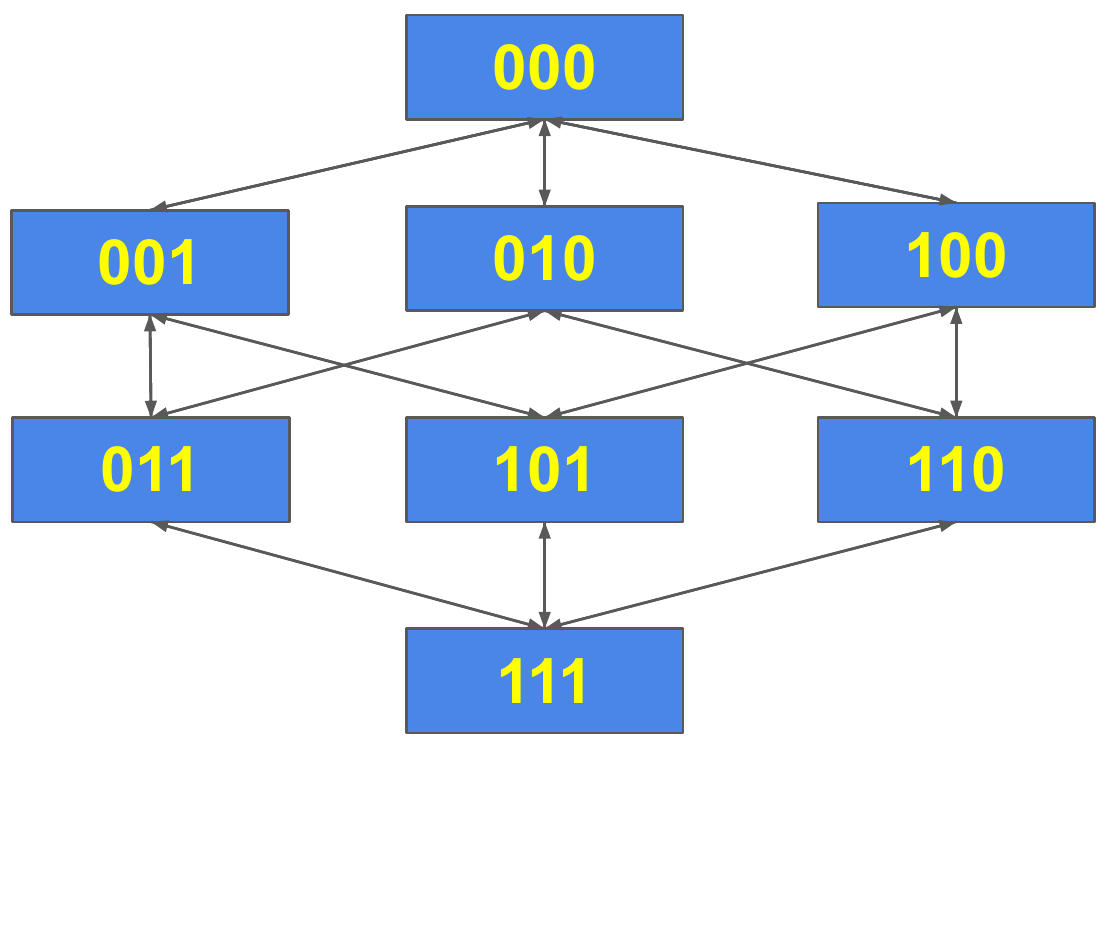}
  \caption{Memristive networks considered in this paper:  (a) $N$ memristors connected in-parallel, and (b) 3-memristor circuit. (c) Transition scheme for a circuit of 2 memristors. (d) Transition scheme for the circuit in (b). (a) and (c) are reprinted with permission from Ref.~\citeonline{dowling2020probabilistic}.}\label{fig:1}
\end{figure}

Consider a circuit composed of $N$ stochastic memristors, any number of ordinary (non-stochastic) resistors, and voltage and current sources. Figure~\ref{fig:1} (a) and (b) shows examples of such circuits. For the sake of simplicity, only binary memristors are considered which means that any of our stochastic devices have two possible states with certain resistances $R=R_{on}$ and $R=R_{off}$. Switching between these states is assumed to be instantaneous, and the switching probability is a definite function of the voltage applied across the memristor.

The circuit state can be specified by an enumeration of the states of all memristors. Let us denote the multi-index $\Theta=(...kji)$, where $i$ (0/1 for the off/on-state) is the state of the first device, $j$ is the state of the second one, and so on. It is clear that at any given moment of time the network state is completely determined by the  multi-index $\Theta$ with $2^N$ possible states.

Due to the probabilistic nature of memristor switching, all network dynamics are unique realizations of the stochastic switching processes. In order to describe the average evolution of a memristive network, we introduce the time-dependent occupation probabilities of circuit states, $p_{\Theta}(t)$. If the stochastic process is Markovian, then the occupation probabilities satisfy the master equation.
Following Ref.~\citeonline{dowling2020probabilistic}, the master equation is written as
\begin{equation}
\frac{\textnormal{d}p_{\Theta}(t)}{\textnormal{d}t}=
\sum\limits_{m=1}^{N}\left(\gamma_{\Theta_m}^mp_{\Theta_m}(t)-\gamma_\Theta^m p_{\Theta}(t) \right)\;,
\label{eq:kin}
\end{equation}
where $\Theta_m$ is the network state obtained from $\Theta$ by flipping the state of $m$-th memristor, $\gamma_\Theta^m$ are the transition rates for $m$-th memristor in the configuration $\Theta$,
%(given by, e.g., Eqs.~(\ref{eq:gamma01}) and (\ref{eq:gamma10})),
 and $\gamma_{\Theta_m}^m$ is defined similarly. Note that the sum of all probabilities is 1.

All information about any particular circuit is contained in the voltage-dependent transition rates, $\gamma^m_\Theta$ and $\gamma^m_{\Theta_m}$, which are calculated for the corresponding circuit states with the use of Kirchhoff's laws. Moreover, to completely determine the occupation probabilities utilizing the master equation (\ref{eq:kin}), the initial conditions for all $p_{\Theta}$ must be specified.

It is convenient to graphically represent all possible transitions for a circuit. For example, the full transition scheme for the case of two memristors is presented in Fig.~\ref{fig:1}(c). The arrows, along with the transition rates, show the direction and frequency of transitions between four possible circuit states, $00$, $01$, $10$, and $11$. In many cases, some of the transitions appear forbidden, i.e. the corresponding transition rates are equal to zero. This leads to a reduced transition scheme and several such examples are considered below. If, additionally, the circuit possesses a symmetry, then
the master equation can be further simplified by reducing the number of independent occupation probabilities. This is allowed since some of the circuit configurations are equivalent and, therefore, the corresponding occupation probabilities coincide.

%\begin{figure}[tb]
%  \centering
%     \includegraphics[height=0.34\columnwidth]{fig1ME}
%  \caption{ Full transition scheme for 2 memristors.} \label{fig:1ME}
%\end{figure}

To illustrate the master equation method, consider the switching dynamics of $N$ identical memristors connected in-parallel to a constant voltage source (see Fig.~\ref{fig:1}(a)). Suppose that only the off- to on-state transitions are allowed, and all memristors are in the off-state at $t=0$. In this highly symmetrical case, the master equation (\ref{eq:kin}) simplifies to $N$ identical equations for one memristor occupation probability
\begin{equation}\label{eq:ap2:1}
  \frac{\textnormal{d}p_{0}(t)}{\textnormal{d}t} = -\gamma_{0}^1p_{0},
\end{equation}
whose solution
\begin{equation}\label{eq:ap2:2}
  p_{0}(t)=e^{-\gamma_{0}^1t}
\end{equation}
yields the probability to find the memristor in the off-state. The probability to find it in the on-state is
\begin{equation}\label{eq:ap2:22}
  p_{1}(t)=1-e^{-\gamma_{0}^1t}.
\end{equation}
The total occupation probability of the circuit is a product of the one memristor occupation probabilities, $p_\Theta(t)=...\cdot p_k(t)p_j(t)p_i(t)$. The probability of complete switching of all $N$ memristors connected in-parallel is equal
\begin{equation}\label{eq:ap2:3}
  p_{1...1}(t)=p_1^N(t)=\left(1-e^{-\gamma_{0}^1t}\right)^N.
\end{equation}

In finding the solution of the master equation, various average characteristics can be calculated for any stochastic memristor in the circuit and for the total circuit (or any of its parts). For example, the average resistance of memristor 1 can be immediately calculated as
\begin{equation}\label{eq:R1}
\langle R_{1}(t) \rangle =R_{off}p_{0}(t)+R_{on} p_{1}(t).
\end{equation}

To demonstrate a calculation for the collective circuit, consider the average network switching time for $N$ identical stochastic memristors connected in-parallel. This time can be evaluated as
\begin{equation}\label{eq:ap2:4}
 \langle T_{\|,N}\rangle=\int\limits_0^\infty t \frac{\textnormal{d}p_{1...1}}{\textnormal{d}t}\textnormal{d}t=
 \frac{1}{\gamma_{0}^1}\left(1+ \frac{1}{2}+ \frac{1}{3}+...+ \frac{1}{N} \right).
\end{equation}

For a more complex example of the master equation approach, we study the truly collective dynamics of a network comprising $N$ identical stochastic memristors connected in-series.
To compare with the circuit of connected in-parallel memristors, we utilize a constant voltage source that is $N$ times greater. Similarly, we suppose that only off- to on-state transitions are allowed, and all memristors are in the off-state at $t=0$.

The master equation for this network can be solved completely analytically and all occupation probabilities can be explicitly evaluated\cite{dowling2020probabilistic}. Here we present the result of calculating the average network switching time for $N$ identical stochastic memristors connected in-series
\begin{equation}
\label{eq:App:sol13a}
  \left< T_N\right>=\sum_{j=0}^{N-1}\frac{1}{(N-j)\gamma_j},
\end{equation}
where $\gamma_j$ is the transition rate of a memristor in the off-state to the on-state in the circuit configuration with $j$ memristors in the on-state.

To compare the switching times (given by Eqs.~(\ref{eq:ap2:4}) and (\ref{eq:App:sol13a})) for the in-parallel and in-series connected memristors, we note that for the transition rates the following inequalities are valid: $\gamma_0=\gamma_0^1$, and $\gamma_j>\gamma_0^1$ for $0<j<N$. This leads to the inequality $\left< T_N\right>< \left< T_{\|,N}\right>$, which means the switching of memristors connected in-series occurs faster than the switching of in-parallel connected ones. Such behaviour can be attributed to the voltage divider effect, where the switching of one memristor in the in-series connected circuit leads to a voltage increase across other memristors accelerating their switching.

Let us conclude this section by noting that the master equation in the form of Eq.~(\ref{eq:kin}) can also be directly applied to the description of the multi-step memristor switching when the binary memristor model is not valid.  Furthermore, by using the Chapman-Kolmogorov equation it is possible to generalize the master equation approach to circuits combining not only stochastic memristors, but also reactive components such as capacitors and inductors~\cite{Slipko21a}.

\section{Circuit dynamics} \label{sec:dynamics}

%%%%%%%%%%%%%%%%%%%%%%%%%%%%%%%%%%%%%%%%%%%%%%%%%%%%%%%%%%%%%
\subsection{SPICE model} \label{sec:SPICE}

While SPICE simulators have been extremely important in the development of deterministic electronic circuits, such as circuits consisting of transistors, resistors, capacitors, inductors, other traditional electronic components as well as emerging components described by deterministic models~\cite{Biolek13a,Benderli09a,Biolek2009-1,rak10a,vourkas2015spice,li2015memristor,garcia2016spice,Biolek16a}, the probabilistic SPICE modeling is an emerging topic~\cite{dowling2020SPICE}.

Here, we consider the dynamics of Fig.~\ref{fig:1}(b) circuit  subjected to a rectangular voltage waveform. In this case, Eq.~(\ref{eq:kin}) is embodied as
\begin{eqnarray}
\frac{\textnormal{d} p_{000}}{\textnormal{d}t} &=&-\left(\gamma_{000}^{1}+\gamma_{000}^{2}+\gamma_{000}^{3}\right)p_{000}+\gamma_{001}^1p_{001}+\gamma_{010}^2p_{010}+\gamma_{100}^3p_{100}\;, \label{syst1a}\\
\frac{\textnormal{d}p_{001}}{\textnormal{d}t}&=&-\gamma_{001}^1p_{001}-(\gamma_{001}^2+\gamma_{001}^3)p_{001}+\gamma_{000}^1p_{000}+\gamma_{011}^2p_{011}+\gamma_{101}^3p_{101}\;,\\
\frac{\textnormal{d}p_{010}}{\textnormal{d}t}&=&-\gamma_{010}^2p_{010}-(\gamma_{010}^1+\gamma_{010}^3)p_{010}+\gamma_{000}^2p_{000}+\gamma_{011}^1p_{011}+\gamma_{110}^3p_{110}\;,\\
\frac{\textnormal{d}p_{100}}{\textnormal{d}t}&=&-\gamma_{100}^3p_{100}-(\gamma_{100}^1+\gamma_{100}^2)p_{100}+\gamma_{000}^3p_{000}+\gamma_{101}^1p_{101}+\gamma_{110}^2p_{110}\;,\\
\frac{\textnormal{d}p_{011}}{\textnormal{d}t}&=&-(\gamma_{011}^1+\gamma_{011}^1)p_{011}-\gamma_{011}^3p_{011}+\gamma_{010}^1p_{010}+\gamma_{001}^2p_{001}+\gamma_{111}^3p_{111}\;,\\
\frac{\textnormal{d}p_{101}}{\textnormal{d}t}&=&-(\gamma_{101}^3+\gamma_{101}^1)p_{101}-\gamma_{101}^2p_{101}+\gamma_{001}^3p_{001}+\gamma_{100}^1p_{100}+\gamma_{111}^2p_{111}\;,\\
\frac{\textnormal{d}p_{110}}{\textnormal{d}t}&=&-(\gamma_{110}^2+\gamma_{110}^3)p_{110}-\gamma_{110}^1p_{110}+\gamma_{100}^2p_{100}+\gamma_{010}^3p_{010}+\gamma_{111}^1p_{111}\;,\\
\frac{\textnormal{d}p_{111}}{\textnormal{d}t}&=&-(\gamma_{111}^1+\gamma_{111}^2+\gamma_{111}^3)p_{111}+\gamma_{011}^3p_{011}+\gamma_{101}^2p_{101}+\gamma_{110}^1p_{110}\label{syst1h}\;,
\end{eqnarray}
where the switching rates depend on the voltage.
Assuming identical parameters for all memristors and deterministic initial state (e.g., $p_{000}(t=0)=1$, $p_{ijk}(t=0)=0$ for $ijk\neq 000$), we introduce $P_0\equiv p_{000}$, $P_1= p_{001}+p_{010}+p_{100}$, $P_2= p_{011}+p_{101}+p_{110}$, $P_3\equiv p_{111}$. Summing up some of the above equations, one then obtains
\begin{eqnarray}
\frac{\textnormal{d} P_{0}}{\textnormal{d}t}&=&-3\gamma_{000}^1P_{0}+\gamma_{001}^1P_{1}\;, \label{eq_3_p0} \\
\frac{\textnormal{d} P_{1}}{\textnormal{d}t}&=&3\gamma_{000}^1P_{0}-\gamma_{001}^1P_{1}-2\gamma_{001}^2P_{1}+2\gamma_{011}^2P_{2}\;,\label{eq_3_p1} \\
\frac{\textnormal{d} P_{2}}{\textnormal{d}t}&=&2\gamma_{001}^2P_{1}-2\gamma_{011}^2P_{2}-\gamma_{011}^3P_{2}+3\gamma_{111}^1P_{3}\;, \label{eq_2_p0}\\
\frac{\textnormal{d} P_{3}}{\textnormal{d}t}&=&\gamma_{011}^3P_{2}-3\gamma_{111}^1P_{3}\;. \label{eq_3_p3}
\end{eqnarray}

\begin{figure}
   \begin{center}
\includegraphics[width=0.95\textwidth]{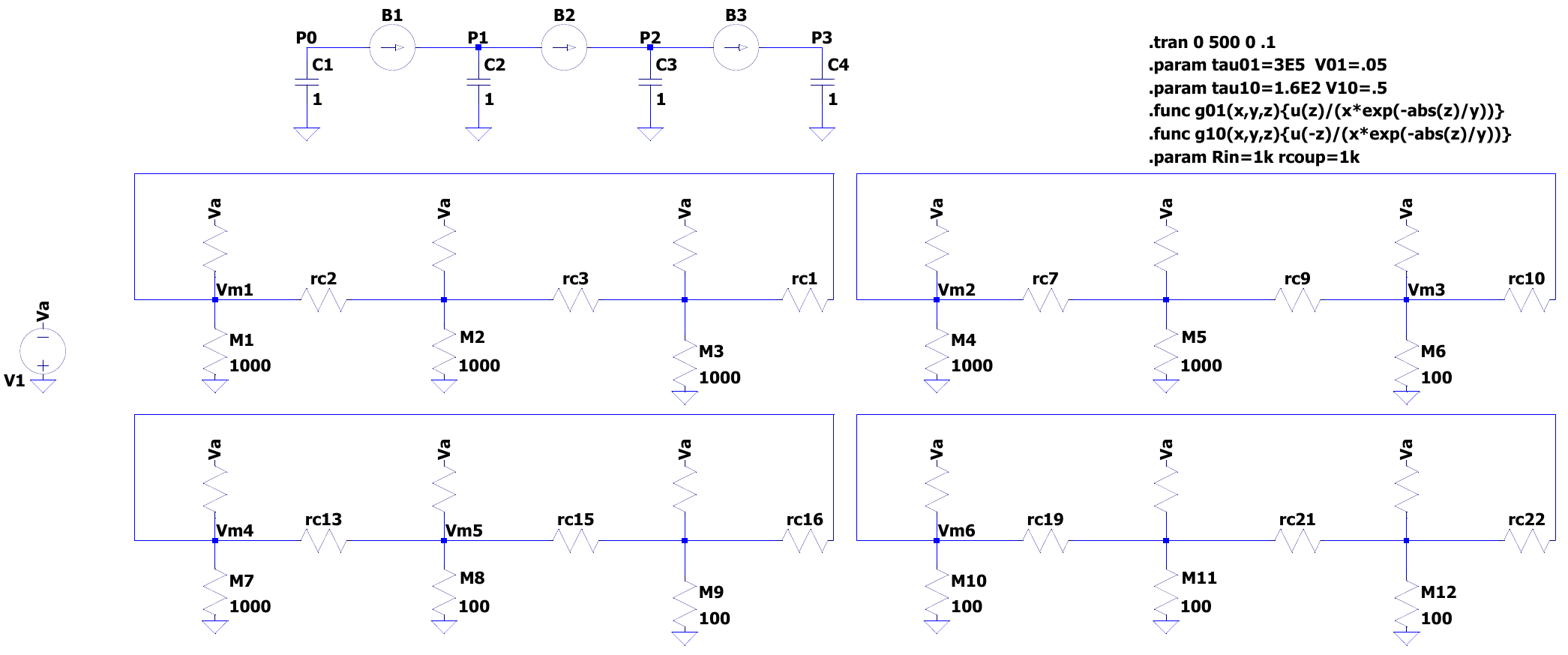}
   \end{center}
   \caption{ \label{fig:2}
SPICE model for the memristive circuit considered (see Fig.~\ref{fig:1}(b)). The rectangular voltage source has a symmetric peak-to-peak voltage of 2~V with a full period of .2s. Each memristor has on and off resistance values of .1~kOhm and 1~kOhm, respectively. Each resistor has a resistance of 1~kOhm.}
   \end{figure}

This system of equations (\ref{eq_3_p0})-(\ref{eq_3_p3}) can be implemented in the SPICE environment following these general rules:
\begin{itemize}
  \item Probabilities of  states are represented by the
  voltage across a 1 Farad capacitor.
  \item Transitions between states are represented by voltage-controlled current sources.
  \item The initial charge of capacitors represents the memristor states at $t=0$.
  \item Several copies of the circuit, differing by the states of memristors, are used to find the voltage across memristors in particular circuit states (used in the calculation of switching rates).
\end{itemize}
Figure \ref{fig:2} gives the SPICE schematic necessary for representing Eqs.~(\ref{eq_3_p0})-(\ref{eq_3_p3}) for the circuit shown in Fig.~\ref{fig:1} (b). Table \ref{tab:0} lists the SPICE code used for each voltage-controlled current source.

In order to account for the voltage-dependent transition rates, $\gamma_{ijk}^m$, 4 copies of the network in each possible resistance configuration are included. Each $Vm_{n}$ in Fig.~\ref{fig:2} gives a voltage across a memristor necessary for describing the evolution of the occupation probabilities. These voltages are used in the calculation of transition (switching) rates.

\begin{table}[h]
\begin{center}
 \begin{tabular}{ | m{1.5em} | m{11.5cm}| }
  \hline
 B1& I=3*g01(tau01,V01,V(Vm1,0))*V(P0)-g10(tau10,V10,V(Vm3,0))*V(P1)  \\
  \hline
B2 & I=2*g01(tau01,V01,V(Vm2,0))*V(P1)-2*g10(tau10,V10,V(Vm5,0))*V(P2)
  \\
  \hline
  B3 & I=g01(tau01,V01,V(Vm4,0))*V(P2)-3*g10(tau10,V10,V(Vm6,0))*V(P3) \\
  \hline
\end{tabular}
\end{center}
\caption{\label{tab:0} SPICE code used to program each voltage-controlled current source for the memristive circuit discussed in this work.}
\end{table}

\begin{figure}
(a) \includegraphics[width=0.45\textwidth]{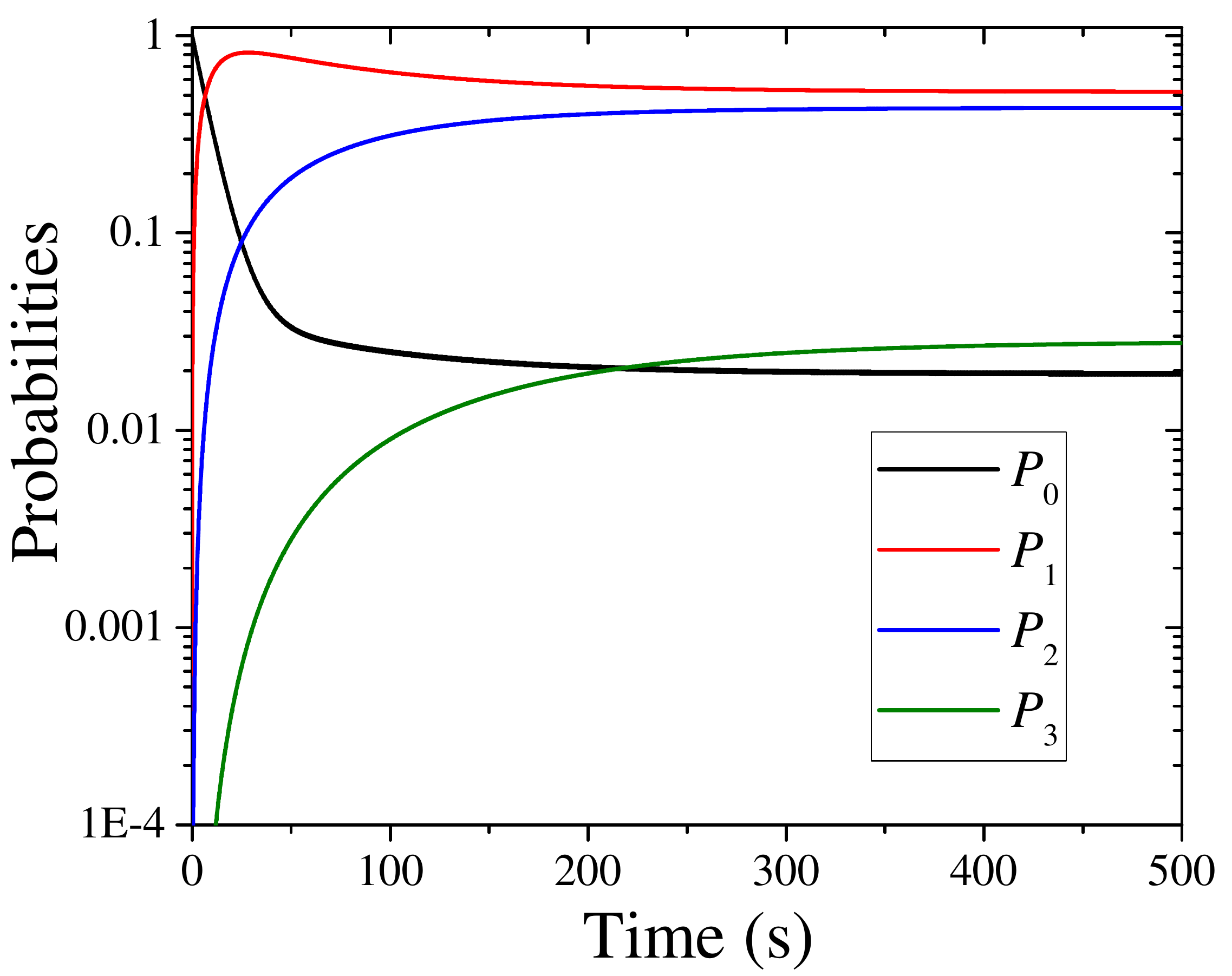} $\;$ (b)\includegraphics[width=0.45\textwidth]{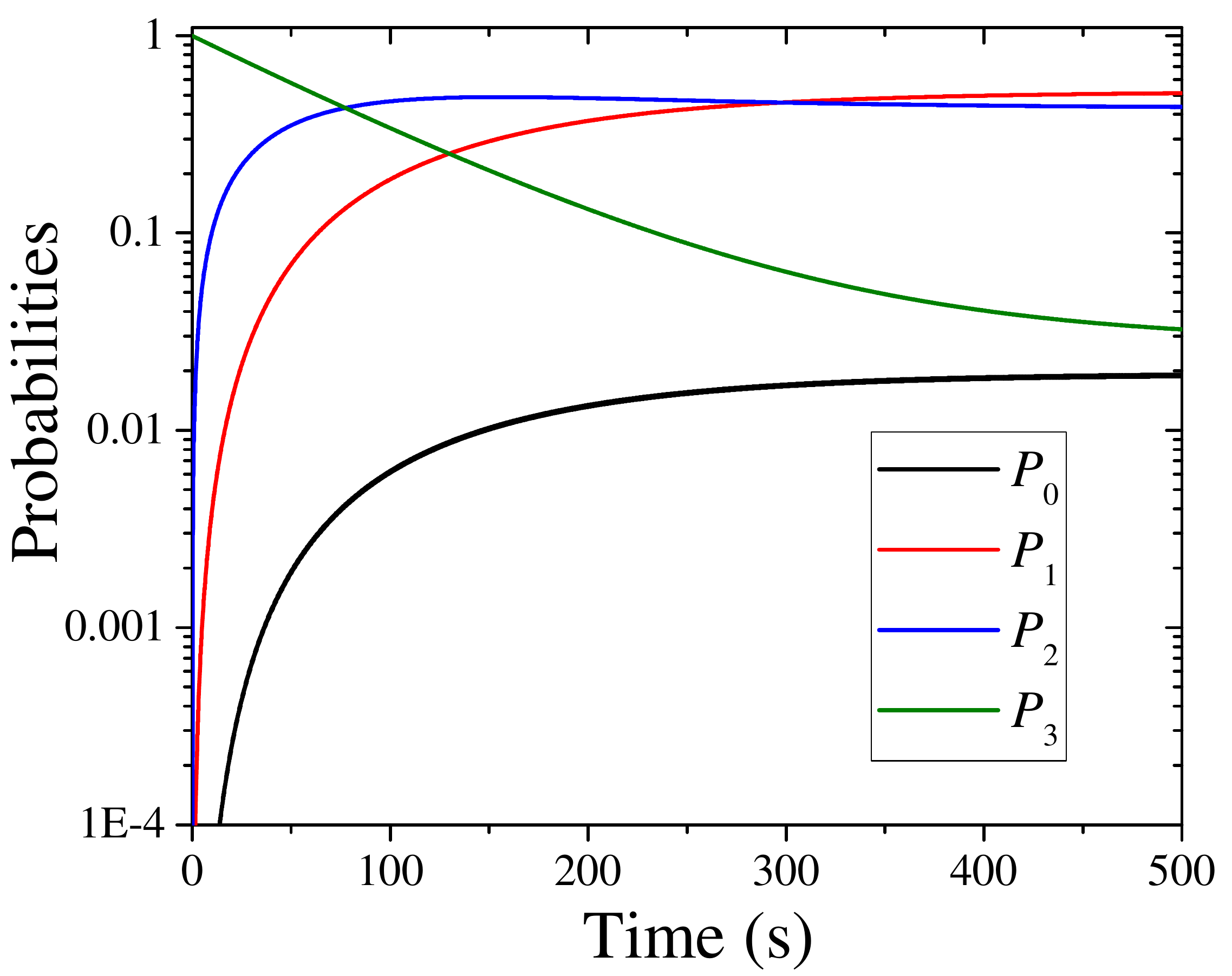}
\caption{Evolution of the occupation probability for each circuit configuration for different initial conditions: (a) $P_0(0)=1$, $P_i(0)=0$ for $i=\{ 1,2,3\}$, and (b) $P_3(0)=1$, $P_i(0)=0$ for $i=\{ 0,1,2\}$. \label{fig:3}}
\end{figure}

\subsection{Numerical results}

%%%%%%%%%%%%%%%%%%%%%%%%%%%%%%%%%%%%%%%%%%%%%%%%%%%%%%%%%%%%%

The circuit presented in Fig.~\ref{fig:1}(b) was simulated not only in SPICE, but using a Monte Carlo scheme as well. Representative SPICE curves are given in Fig.~\ref{fig:3} that show the dynamics of probabilities for two different cases of initial conditions (all memristors are initially either in the off- or on-state). We emphasize the development of a steady-state in time, that does not show any significant periodic fluctuations. It is explained by small changes of probabilities in each half-period of the voltage.

Next we briefly discuss the calculation of probabilities in a Monte Carlo simulation. The circuit is first initialized and the first set of voltages across each memristor are calculated through Kirchoff’s Laws. Those voltages are used to generate a switching time found through the inverse of the normal transition rates scaled by a randomized log value for each memristor. The fastest switching time is selected and compared to the remaining time in the current voltage period. If there is sufficient time, the memristor will change states and the remaining voltage period will be reduced by the switching time. If the voltage period is shorter than the switching time, the voltage will flip to its other value and the period will be reset. New memristor voltages are calculated and the simulation continues until sufficient time has passed the network to reach its steady-state.

Each simulation used circuit and memristor parameters given in Fig.~\ref{fig:2}. The SPICE simulation has a length of 10,000 seconds and the Monte Carlo simulation has a duration of 1,000,000 seconds with the first 1,000 seconds ignored in order for the circuit to reach its steady-state before recording data. Both simulations utilize a time-step of .1 seconds. The SPICE simulation was done using LTspice XVII. Table~\ref{tab:1} compares steady-state values of circuit states found for both SPICE and Monte Carlo simulations using the above parameters. The convergence of these two methods increases with modeling size.

\begin{table}[h]
\begin{center}
\begin{tabular}{  | m{2cm} | m{2cm} | m{2cm}| m{2cm}|}
  \hline
  Probability& MC& SPICE&Difference  \\
  \hline
$P_0$ &0.0198477&0.01953948&1.565080\%
  \\
  \hline
  $P_1/3$ &0.1726743& 0.17344830&0.447240\%  \\
  \hline
$P_2/3$& 0.1444423&0.14397540&0.323767\%\\
\hline
$P_3$& 0.0288020&0.02818918&2.150580\%\\
\hline
\end{tabular}
\end{center}
\caption{\label{tab:1} Steady state probability values: Comparison of SPICE and Monte Carlo simulations.}
\end{table}

\subsection{Circuit attractor} \label{sec:attractor}

The same steady state found in calculations presented in Fig.~\ref{fig:3}(a) and (b) can be interpreted as the circuit attractor that is analogous with attractors in deterministic memristor circuits~\cite{Pershin2019a}. Previously, two of us (YVP and VAS) predicted the possibility of attractor states in deterministic memristor circuits driven by alternating polarity pulses. It was shown that, on average, the set of internal state variables of all memristors may converge to a single point in phase space. In circuits of stochastic memristors considered here, attractors emerge in the space of state probabilities (such as $P_i(t)$).

Assume that a circuit containing stochastic memristors is driven by short alternating polarity pulses whose amplitude and duration are $V_\pm$ and $\tau_\pm$, respectively. If an attractor state exists, after a long period of time the circuit state oscillates about an equilibrium state. In this situation, the change in the circuit state by a positive pulse is fully compensated by the subsequent negative pulse. For Eqs.~(\ref{eq_3_p0})-(\ref{eq_3_p3}), this condition can be presented by the linear system
\begin{eqnarray}
-3\gamma_{000}^1(V_+)P_{0}\tau_++\gamma_{001}^1(V_-)P_{1}\tau_-&=&0\;, \label{1eq_3_p0} \\
3\gamma_{000}^1(V_+)P_{0}\tau_+-\gamma_{001}^1(V_-)P_{1}\tau_--2\gamma_{001}^2(V_+)P_{1}\tau_++2\gamma_{011}^2(V_-)P_{2}\tau_-&=&0\;,\label{1eq_3_p1} \\
2\gamma_{001}^2P_{1}\tau_+-2\gamma_{011}^2(V_-)P_{2}\tau_--\gamma_{011}^3(V_+)P_{2}\tau_++3\gamma_{111}^1(V_-)P_{3}\tau_-&=&0\;, \label{1eq_2_p0}\\
\gamma_{011}^3(V_+)P_{2}\tau_+-3\gamma_{111}^1(V_-)P_{3}\tau_-&=&0\; \label{1eq_3_p3}
\end{eqnarray}
with unknown $P_i$ satisfying the normalization condition $P_0+P_1+P_2+P_3=1$. In the above equations, the voltage in parenthesis represents the applied voltage at the time the voltage across the transitioning memristor is calculated. In principle, one can recognize just three independent equations in the system (\ref{1eq_3_p0})-(\ref{1eq_3_p3}) that are
\begin{eqnarray}
-a_1P_{0}+b_1P_{1}&=&0\;, \label{2eq_3_p0} \\
-a_2P_{1}+b_2P_{2}&=&0\;,\label{2eq_3_p1} \\
-a_3P_{2}+b_3P_{3}&=&0\; \label{2eq_3_p2},
\end{eqnarray}
where $a_1=3\gamma_{000}^1(V_+)\tau_+$, $b_1=\gamma_{001}^1(V_-)\tau_-$,
$a_2=2\gamma_{001}^2(V_+)\tau_+$, $b_2=2\gamma_{011}^2(V_-)\tau_-$,
$a_3=\gamma_{011}^3(V_+)\tau_+$, and $b_3=3\gamma_{111}^1(V_-)\tau_-$.
These equations represent the condition of zero probability flow among the states within each period, and can be easily solved analytically.

\begin{figure}
\centering \includegraphics[width=0.45\textwidth]{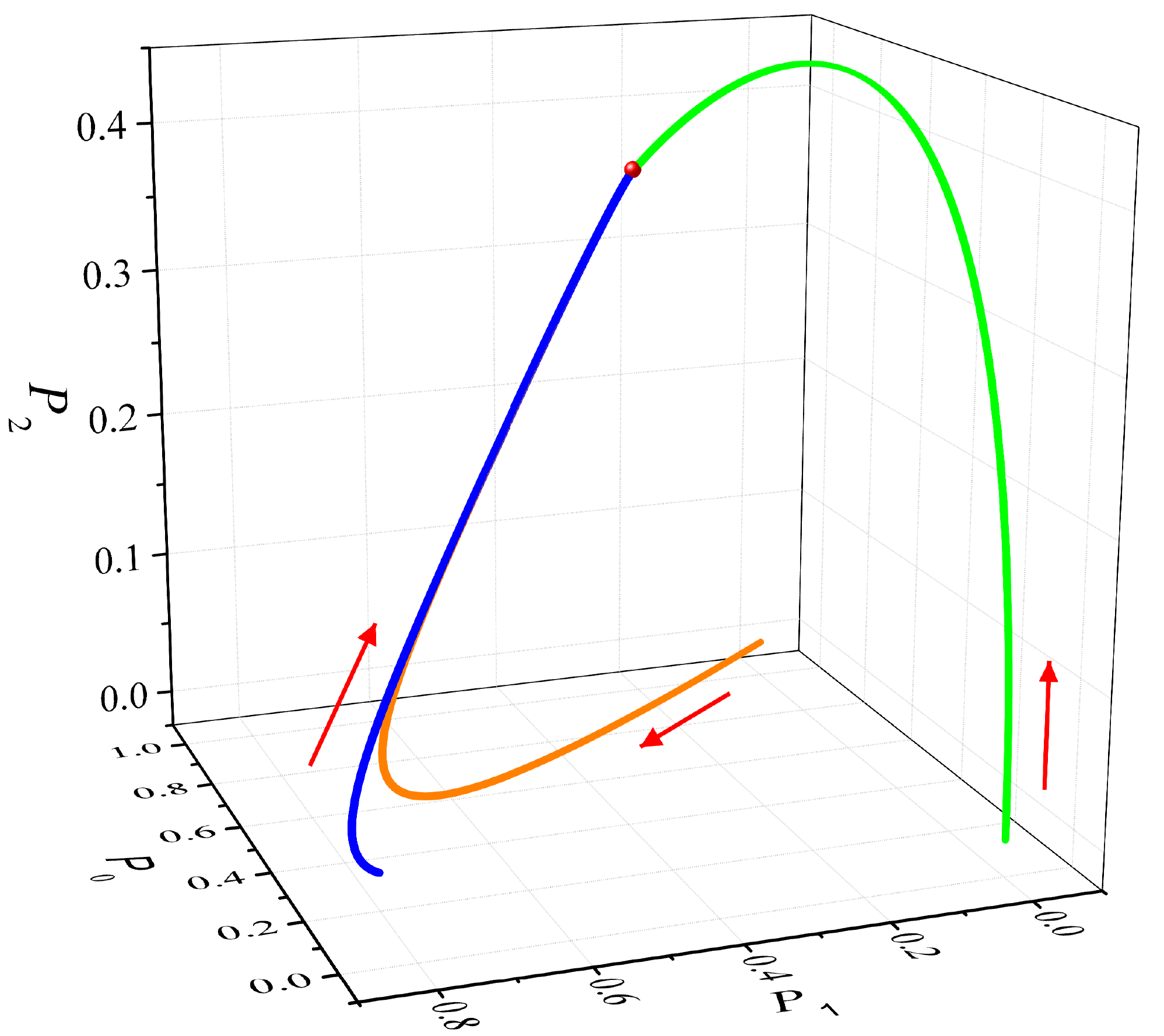}
\caption{Probabilistic circuit attractor: convergence of trajectories corresponding to distinct initial conditions to a single point. Two curves in this plot represent the calculations in Fig.~\ref{fig:3}, while the third curve was obtained for a probabilistic initial state $P_0(0)=0.2$, $P_1(0)=0.8$, $P_{3(4)}(0)=0$. The arrows show the direction of evolution.  \label{fig:4}}
\end{figure}

The solution of
Eqs.~(\ref{2eq_3_p0})-(\ref{2eq_3_p2})  supplemented by the normalization condition (below Eq.~(\ref{1eq_3_p3})) can be presented as
\begin{equation}
P_0=\frac{1}{1+\frac{a_1}{b_1}+\frac{a_1a_2}{b_1b_2}+\frac{a_1a_2a_3}{b_1b_2b_3}},\;\; P_1=\frac{a_1}{b_1}P_0,\;\;P_2=\frac{a_1a_2}{b_1b_2}P_0,\;\;P_3=\frac{a_1a_2a_3}{b_1b_2b_3}P_0.  \label{Eq:stablepoint}
\end{equation}
Using the circuit parameters from Fig.~\ref{fig:2}, $V_+=-V_-=1$~V, $\tau_+=\tau_-=.1$~s, one can find $P_0=0.0189997$, $P_1/3=0.172468$, $P_2/3=0.144958$, and $P_3=0.0287214$, which are in an excellent agreement with the values in Table~\ref{tab:1}.  Fig.~\ref{fig:4}  shows the convergence of trajectories to a single point of phase space -- the circuit attractor, which is represented by the red point. Fig.~\ref{fig:4} was generated using the model in Fig.~\ref{fig:2} and using the values given below Eq.~(\ref{Eq:stablepoint}).

The question about the existence and uniqueness of an attractor in periodically-driven circuits with stochastic memristors arises naturally. In fact, under appropriate conditions (such as driving by short alternating polarity pulses), the master equation can be averaged over the period of pulses and its right-hand side can be rewritten with some constant transition rates. From the theory of such equations, it is known that under certain conditions~\cite{vanCampen} there exists a unique steady-state solution, which is asymptotic for all possible initial conditions. The attractor discussed above is an example of such a steady-state solution.

\section{Conclusion}

The contribution of this conference paper is twofold. First,  to better represent circuits of stochastic memristors in SPICE, we have optimized SPICE models such that capacitors used to store state probabilities are now directly coupled to each other through voltage-controlled current sources. Compared to Ref.~\citeonline{dowling2020SPICE}, this has simplified the equations of current sources and reduced the number of current sources required by one.
Second, we demonstrated a transition into an attractor state in a circuit of stochastic binary memristors periodically driven by alternating polarity pulses, and derived analytically the attractor location. We hope that this work will be useful for scientists and students with interest in the area of emerging memory devices.

%%%%%%%%%%%%%%%%%%%%%%%%%%%%%%%%%%%%%%%%%%%%%%%%%%%%%%%%%%%%%
%\acknowledgments     %>>>> equivalent to \section*{ACKNOWLEDGMENTS}

%This unnumbered section is used to identify those who have aided the authors in understanding or accomplishing the work presented and to acknowledge sources of funding.

%%%%%%%%%%%%%%%%%%%%%%%%%%%%%%%%%%%%%%%%%%%%%%%%%%%%%%%%%%%%%
%%%%% References %%%%%

\bibliography{virus,memcapacitor}   %>>>> bibliography data in report.bib
\bibliographystyle{spiebib}   %>>>> makes bibtex use spiebib.bst

\end{document}